\documentclass[english,,aps,prl,twocolumn,superscriptaddress,hyperref]{revtex4}
\usepackage{helvet}
\usepackage[T1]{fontenc}
\usepackage[latin1]{inputenc}
\usepackage{graphicx}
\usepackage{amssymb}

\makeatletter

\makeatletter

\makeatletter

\makeatletter

\makeatletter

\usepackage{latexsym}

\usepackage{comment}

\makeatother

\makeatother

\makeatother

\makeatother

\usepackage{babel}
\makeatother
\begin{document}

\title{Universal thermodynamics of strongly interacting Fermi gases}

\author{Hui Hu}

\affiliation{Department of Physics, Renmin University of China, Beijing 100872,China}

\affiliation{ARC Centre of Excellence for Quantum-Atom Optics, Department of Physics,
University of Queensland, Brisbane, Queensland 4072, Australia}

\author{Peter D. Drummond}

\email{drummond@physics.uq.edu.au}

\affiliation{ARC Centre of Excellence for Quantum-Atom Optics, Department of Physics,
University of Queensland, Brisbane, Queensland 4072, Australia}

\author{Xia-Ji Liu}

\affiliation{ARC Centre of Excellence for Quantum-Atom Optics, Department of Physics,
University of Queensland, Brisbane, Queensland 4072, Australia}

\date{\today}

\maketitle
\textbf{Strongly interacting Fermi gases are of great current interest.
Not only are fermions the most common particles in the universe, but
they are also thought to have a universal thermodynamic behavior for
strong interactions~\cite{heiselberg,carlson,ho}. Recent experiments
on ultra-cold Fermi gases provide an unprecedented opportunity to
test universality in the laboratory~\cite{hara,thomas,partridge,stewart,luo}.
In principle this allows - for example - the interior properties of
hot, dense neutron stars to be investigated on earth. Here we carry
out a detailed test of this prediction. We analyze results from three
ultra-cold fermion experiments involving two completely distinct atomic
species in different kinds of atomic trap environments~\cite{partridge,stewart,luo}.
The data is compared with the predictions of a recent strong interaction
theory~\cite{hldepl,hldpra}. Excellent agreement is obtained, with
no adjustable parameters. By extrapolating to zero temperature, we
show that the experimental measurements yield a many-body parameter}
$\mathbf{\beta\simeq-0.59\pm.07}$, \textbf{describing the universal
energy of strongly interacting Fermi gases.}

Experiments on ultra-cold Fermi gases at micro-Kelvin temperatures
are revolutionizing many areas of physics. Their exceptional simplicity
allows tests of many-body theory in areas long thought to be inaccessible.
The ability to widely tune the effective interaction between fermions
via a broad Feshbach resonance in gases of $^{6}$Li and $^{40}$K
has allowed resonance models proposed in high-Tc superconductivity
theory~\cite{FriedbergLee} to be implemented with fermionic ultra-cold
atoms~\cite{KheruntsDrummond}. Fortunately, the Pauli exclusion
principle stabilizes the resulting excited molecular states against
collisional damping~\cite{Petrov}. This has permitted the experimental
observation of the smooth evolution of the Fermi gas from the attractive
regime of Bardeen-Cooper-Schrieffer (BCS) superfluidity through to
a regime where molecules form a Bose-Einstein condensate (BEC)~\cite{leggett,regal,bourdel,bartenstein,zwierlein,stoof}.
On the cusp of this BCS-BEC crossover, there is a strongly interacting
regime --- the so-called unitarity limit~\cite{ho}, which leaves
the inter-atomic distance as the only relevant length scale.

At this point, the gas is expected to exhibit a universal thermodynamic
behaviour, independent of any microscopic details of the underlying
interactions. The ground state energy $E_{0}$ of a homogeneous gas
at zero temperature should be proportional to the free Fermi energy,
$E_{F}$. Thus, $E_{0}=E_{F}(1+\beta)$, where $\beta$ is a universal
many-body parameter. Substantial experimental efforts have been carried
out to verify the existence of universality~\cite{hara,thomas},
though so far there has been no conclusive confirmation. This is mainly
due to the lack of reliable thermometry in the strongly interacting
regime~\cite{hldpra}, leading to an uncertainty in the finite temperature
corrections.

The universal parameters estimated from energies at the lowest accessible
(but unknown) temperature, range from $\beta=-.68\pm0.1$ to $\beta=-.54\pm0.05$
~\cite{partridge,luo,bartenstein,bourdel}. Since these are not at
zero temperature, there is an unknown correction factor required to
obtain the ground state energy, and hence the true value of $\beta.$
There is a similar range of estimated theoretical values, though a
more precise value of $\beta=-0.58\pm0.01$ was recently obtained
from zero temperature quantum Monte Carlo simulations~\cite{astrakharchik}.

This situation has dramatically improved in the most recent thermodynamic
measurements on strongly interacting Fermi gases of $^{40}$K and
$^{6}$Li atoms~\cite{stewart,luo}, which allow accurate estimates
of the energy in the universal regime from the fermionic cloud size.
In experiments on $^{40}$K carried out at JILA~\cite{stewart},
an adiabatic magnetic field sweep is used to compare measurements
in the strongly interacting and weakly interacting regimes, so that
the non-interacting temperature is also known from the cloud size
after the sweep. An important conceptual advance of the Duke group~\cite{thomas}
who use $^{6}$Li, was the realization that this gives a model-independent
measurement of the entropy in the strongly-interacting regime, thus
allowing a precision test~\cite{luo} of theoretical predictions
of universal thermodynamics. A different approach at Rice~\cite{partridge},
also with $^{6}$Li, makes use of the detailed density distribution
to estimate temperature and entropy.

These ground-breaking investigations provide measurements accurate
to the level of a few percent, which is an exceptional accuracy in
this challenging field of ultra-low temperature physics.

\begin{figure}
\centering \includegraphics[width=88mm]{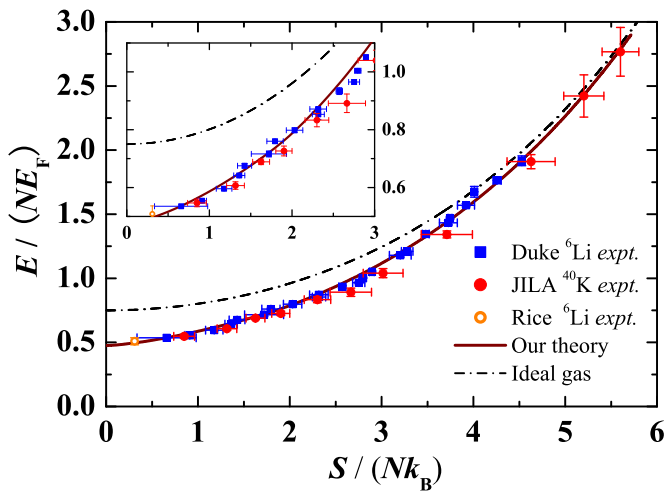}

\caption{(Color online) \textbf{Illustration of the universal thermodynamics
of a strongly interacting Fermi gas}. Comparison between theoretical
predictions and experimental measurements on the entropy-energy relation
of strongly interacting Fermi gases in a harmonic trap. Inset highlights
the low entropy region. The energy is in units of the Fermi energy,
which is the highest single-particle energy level of a non-interacting
gas in the same trap with the same number of fermions, $N$. Brown
solid curve is our theoretical prediction, and the black dash-dotted
curve is the ideal gas energy. The experimental data is from Refs.~\cite{partridge,stewart}
and~\cite{luo}, on $^{6}Li$ and $^{40}K$ fermionic atoms, with
error-bars taken from the experimental papers using appropriate conversions.
All results agree with a single, universal theoretical curve.}

\label{fig-universal} 
\end{figure}

In this Letter, we re-analyze all the available experimental data
from these three laboratories, thus obtaining the measured entropy-energy
relation of two completely different strongly interacting trapped
Fermi gases. We compare this directly with a single universal theoretical
prediction. We use a diagrammatic approach based on functional path-integrals~\cite{engelbrecht,hldepl}
together with the local density approximation to treat the inhomogeneous
trap. Below the superfluid transition, our calculations include pairing
fluctuations, which are important in the strongly interacting regime,
due to the onset of molecule formation. This approach is described
in detail elsewhere~\cite{hldepl}. Above threshold we use the well-known
Nozi\`{e}res-Schmidt-Rink (NSR) theory~\cite{nsr,ohashi,lhepl}.
We demonstrate a quantitative test of these thermodynamic predictions
by comparing experimental results on both types of atom with a single
theoretical curve, as shown in Fig (\ref{fig-universal}). There are
no adjustable parameters, so this provides strong evidence for universality.

Further, by using power-law extrapolation to estimate finite temperature
corrections, we are able to show that all the experimental data give
rise to a universal many-body coefficient of $\beta\simeq-0.59\pm0.07$.
This is in excellent agreement both with the Monte Carlo results~\cite{astrakharchik},
$\beta=-0.58\pm0.01$, and our earlier theoretical prediction~\cite{hldepl}
of $\beta\approx-0.599$.

We summarize the experimental procedures, as typified by the JILA
work using $^{40}$K atomic gases~\cite{stewart}. Here, the strongly
interacting gas is prepared in a harmonic trap at the Feshbach resonance,
and the potential energy is measured from the observed radius. Next,
the magnetic field is swept adiabatically to a zero scattering length
field, and the potential energy at this field is again measured, which
gives the non-interacting temperature. From this data, we obtain the
entropy of the interacting gas~\cite{stewart}, since the energy-entropy
relation of a non-interacting Fermi gas is known. The total energy
is also obtained, as it is twice the potential energy at the resonance,
owing to the virial theorem~\cite{thomas}. Fig. 2a gives the predicted
potential energy ratio as a function of the non-interacting temperature
in the presence of a harmonic trap, together with the experimental
measurements. The observed reduction of the potential energy in the
strongly interacting regime is theoretically reproduced. Converting
the non-interacting temperature and the potential energy into the
total entropy and energy, respectively, we obtain the entropy-energy
relation for $^{40}$K gas in Fig. 2b. We find an excellent agreement
between the experimental data and theoretical predictions below threshold.
There is a small discrepancy just above the critical temperature of
$(T/T_{F})^{0}\approx0.25$ or the critical entropy of $S_{c}\simeq2.2Nk_{B}$,
where we expect that the above threshold NSR theory may be less reliable.
This effect is clearly visible in Fig. 2a, which gives the original
experimental measurements. The nonlinear transformation used to obtain
the entropy-energy relation in Fig. 2b means that conventional rectangular
error-bars give only a qualitative indication of the uncertainties
in this figure.

\begin{figure}
\centering \includegraphics[width=88mm]{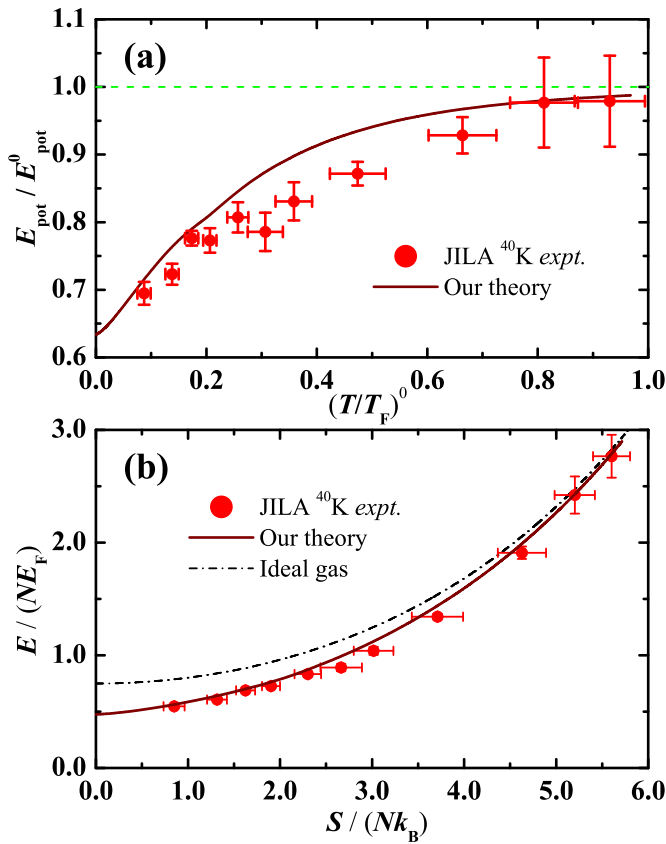}

\caption{(Color online) \textbf{Analysis of the $^{40}$K experiment in JILA}.
\textbf{a}, Experimental data on the potential energy $E_{pot}$ normalized
to that in the non-interacting regime $E_{pot}^{0}$ for a $^{40}$K
gas at unitarity, as a function of the non-interacting gas temperature
$(T/T_{F})^{0}$. It converges to unity as expected (dashed green
horizontal line). The experimental data are compared with our theoretical
prediction for a harmonically trapped, strongly interacting Fermi
gas. \textbf{b}, Comparison between theory and experiment on the entropy
dependence of the energy for a unitarity gas. The experimental entropy
is calculated from the non-interacting temperature, with error-bars
converted from the experimental data. The energy is double the potential
energy $E_{pot}$ due to the virial theorem~\cite{thomas}.}

\label{fig2} 
\end{figure}

Next, we discuss in greater detail the $^{6}$Li experiment in Duke~\cite{luo},
which has greater accuracy but involves some extra corrections due
to the anharmonic trap used, and residual interactions in the weakly
interacting cloud. The strongly interacting Fermi gas of $N=1.3(2)\times10^{5}$
atoms is prepared in a shallow Gaussian (anharmonic) trap $V(\mathbf{r})=U_{0}\{1-\exp[-m(\omega_{\perp}^{2}\rho^{2}+\omega_{z}^{2}z^{2})/(2U_{0})]\}$
at a magnetic field $B=840$ G, slightly above the resonance position
$B_{0}=834$ G. The coupling constant $k_{F}a=-30.0$, where $k_{F}$
is the Fermi wave vector and $a$ is the \textit{s}-wave inter-atomic
scattering length, is sufficient large to ensure the onset of the
universal thermodynamic behavior. Experimentally, the entropy of the
gas is measured by an adiabatic passage to a weak interacting field
$B=1200$ G, where $k_{F}a=-0.75$ and the entropy and temperature
is known from the cloud size after the sweep. The energy $E$ is determined
model independently from the mean square radius of the strongly interacting
fermion cloud $\left\langle z^{2}\right\rangle _{840}$ measured at
$840$ G, according to the virial theorem~\cite{thomas,luo}, \begin{equation}
\frac{E}{NE_{F}}=\frac{\left\langle z^{2}\right\rangle _{840}}{z_{F}^{2}}\left(1-\kappa\right),\end{equation}
 where $E_{F}=(3N\omega_{\perp}^{2}\omega_{z})^{1/3}=k_{B}T_{F}$
is the Fermi energy for an ideal harmonically trapped gas at the trap
center, and $z_{F}^{2}$ is defined by $3m\omega_{z}^{2}z_{F}^{2}\equiv E_{F}$.
The correction factor $1-\kappa$ accounts for the anharmonicity in
the shallow trapping potential $U_{0}\simeq10E_{F}$.

Fig. 3a shows the bare experimental data on the ratio of the mean
square axial cloud size at $1200$ G to that at $840$ G, as a function
of the energy at $840$ G, as compared to the theoretical simulations
with the same realistic parameters, except that we use a resonance
field $B_{0}$ for the strongly interacting gas. Good agreement is
found, with no free parameters. As before, there is a small discrepancy
between the raw data and theoretical predictions, just above the critical
energy. We have recalculated the entropy corrections due to residual
interactions in the $1200$G cloud to improve the accuracy at the
$1\%$ level, by using an above-threshold NSR theory. Calibration
of the entropy from the measured mean square axial cloud size at $1200$
G using the theoretically predicted dependence of the entropy on the
size (inset in Fig. 3b) leads to the comparison for the entropy-energy
relation, as shown in Fig. 3b. The agreement is even more impressive.

\begin{figure}
\centering \includegraphics[width=88mm]{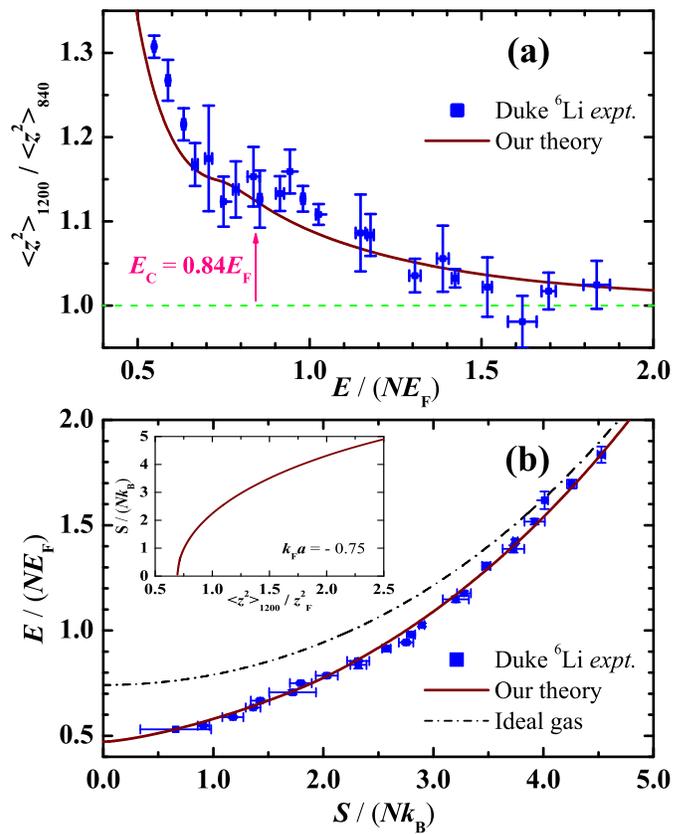}

\caption{(Color online) \textbf{Analysis of the $^{6}$Li experiment in Duke}.
\textbf{a}, Experimental data on the ratio of the mean square cloud
size at 1200 G, $\left\langle z^{2}\right\rangle _{1200}$ to that
at 840 G, $\left\langle z^{2}\right\rangle _{840}$ for a $^{6}$Li
gas, as a function of total energy, are compared to the theoretical
simulations. The data is obtained by an adiabatic passage from a strongly
interacting field $B=840$ G to a weakly interacting field $B=1200$
G. The total energy is measured at $840$ G by utilizing the virial
theorem, and is normalized with respect to the Fermi energy of a non-interacting
gas. The theoretical curve are calculated under the same procedure
and parameters, except that the starting field is at the Feshbach
resonance $B=834$ G. As shown by the dashed green horizontal line,
the ratio converges to unity at high energy as the gas becomes more
ideal. The arrow points to the theoretically predicted energy at transition
point. \textbf{b}, Experimental data on the entropy dependence of
the energy for a strongly interacting gas at $840$ G, compared to
the predictions from our strong interaction theory. The result for
an ideal gas is also plotted. The entropy and error-bars of the gas
are calibrated from the experimental mean square axial cloud size
$\left\langle z^{2}\right\rangle _{1200}$, using our theoretical
dependence of the entropy on the cloud size (as shown in the inset),
which should be extremely accurate in the weakly interacting regime.}

\label{fig3} 
\end{figure}

We can now describe the procedures used to obtain Fig (1), which illustrates
the universal thermodynamic behavior of a strongly interacting Fermi
gas. Here we have plotted all the measured data in a single figure,
and compared them with our prediction for the entropy dependence of
the energy of a harmonically trapped, strongly interacting Fermi gas,
as well as that of an ideal Fermi gas. The slight shift of the experimental
data in the Duke experiment due to the anharmonicity of the trap has
been corrected, by subtracting the (small) theoretical difference
between a shallow Gaussian trap and a harmonic trap for the energy
at the same entropy. We plot also a single data point from the $^{6}$Li
experiment in Rice~\cite{partridge} for the energy at their lowest
entropy. The agreement between theory and experiment is excellent
for almost all the measured data. Exactly the same theory is used
in all cases, with results from three different laboratories~\cite{partridge,stewart,luo}.
The universal thermodynamics of a strongly interacting Fermi gas is
strikingly demonstrated, independently of which atomic species we
compare with.

Just above the critical entropy $S_{c}\simeq2.2Nk_{B}$, for the superfluid-normal
fluid phase transition, there is a suggestion of a discrepancy between
theoretical predictions and these precise measurements. At this point
the above-threshold NSR theory is least accurate~\cite{hldpra}.
The data here may even indicate a first order superfluid phase transition.
We note that the exact order of phase transition for a strongly interacting
Fermi gas is still an open question~\cite{sc06}, and merits further
investigation.

A key feature of current ultra-cold Fermi gas experiments is that
the lowest attainable entropy is around $S=0.7Nk_{B}$, which corresponds
to a temperature of $0.10-0.15T_{F}$ at unitarity. This nonzero entropy
or temperature affects the precise determination of the universal
many-body parameter $\beta$. To remove the temperature dependence,
we assume that in the low entropy regime (below the phase transition),
there is a power law dependence of the energy on the entropy: $E-E_{0}\propto S^{\alpha}$,
as anticipated theoretically. For non-interacting Fermi and Bose gases,
the power law exponents are $2$ and $4/3$, respectively. The thermodynamics
of a unitary gas should lie between an ideal Fermi gas and an ideal
Bose-Einstein condensate. The fitting procedure leads to $E_{0}/(NE_{F})=0.48\pm0.03$
and $E_{0}/(NE_{F})=0.48\pm0.04$, for the Duke and JILA experiments,
respectively, with a similar power law exponent $\alpha=1.7\pm0.4$.
Here the error bar accounts for the fitting error only. Using the
relation $E_{0}/N=(3/4)(1+\beta)^{1/2}E_{F}$ for a harmonic trap~\cite{hara},
this gives rise to $\beta\simeq-0.59\pm0.07$: which agrees fairly
well with the most accurate quantum Monte Carlo simulations~\cite{astrakharchik},
$\beta=-0.58\pm0.01$, and our theoretical predictions~\cite{hldepl},
$\beta\approx-0.599$. Our theoretical power-law prediction is $\alpha=1.5$
in the LDA regime, which also agrees with experiment.


\begin{acknowledgments}
We are extremely grateful to J. E. Thomas and B. Clancy for many helpful
discussions, and for sharing their data before publication. We also
thank D. S. Jin and J. T. Stewart for communications on their data,
and R. G. Hulet et. al. for explaining their temperature measurements.
This work was supported by an Australian Research Council Center of
Excellence grant, the National Natural Science Foundation of China
Grant No. NSFC-10574080, and the National Fundamental Research Program
Grant No. 2006CB921404 and 2006CB921306. Correspondence and requests
for materials should be addressed to P. D. D. 
\end{acknowledgments}

\textbf{\textit{Methods.}} --- We briefly explain our analytic theory.
This is an approximate method using perturbation theory summed to
all orders, since no exact results are known. In the homogeneous gas
case, it relies on the many-body $T$-matrix approximation to account
for the effects of collective Bogoliubov-Anderson modes, and extends
the standard NSR approach to the broken-symmetry state~\cite{hldepl}.
This amounts to considering the contributions of Gaussian fluctuations
around the mean-field saddle point to the thermodynamic potential
(with Nambu notation), \begin{equation}
\delta\Omega=\frac{1}{2}\sum_{Q}\ln\det\left[\begin{array}{cc}
\chi_{11}\left(Q\right) & \chi_{12}\left(Q\right)\\
\chi_{12}\left(Q\right) & \chi_{11}\left({\bf -}Q\right)\end{array}\right],\end{equation}
 where \begin{eqnarray*}
\chi_{11}(Q) & = & \frac{m}{4\pi\hbar^{2}a}+\sum\nolimits _{K}{\cal G}_{11}(Q-K){\cal G}_{11}(K)-\sum_{\mathbf{k}}\frac{1}{2\epsilon_{\mathbf{k}}},\\
\chi_{12}(Q) & = & \sum\nolimits _{K}{\cal G}_{12}(Q-K){\cal G}_{12}(K),\end{eqnarray*}
 are respectively the diagonal and off-diagonal parts of the pair
propagator. Here, $Q=(\mathbf{q},i\nu_{n})$, $K=(\mathbf{k},i\omega_{m})$,
and $\sum_{K}=$ $k_{B}T\sum_{m}\sum_{\mathbf{k}}$ ($\mathbf{q}$
and $\mathbf{k}$ are wave vectors, $\nu_{n}$ and $\omega_{m}$ bosonic
and fermionic Matsubara frequencies, respectively), $m$ is the fermion
mass, $T$ the temperature, and $\epsilon_{\mathbf{k}}=\hbar^{2}\mathbf{k}^{2}/2m$,
${\cal G}_{11}$ and ${\cal G}_{12}$ are BCS Green's functions with
a variational order parameter $\Delta$. Together with the mean-field
contribution \begin{equation}
\Omega_{0}=\sum_{\mathbf{k}}\left[\epsilon_{\mathbf{k}}-\mu+\frac{\Delta^{2}}{2\epsilon_{\mathbf{k}}}+2k_{B}Tf(-E_{\mathbf{k}})\right]-\frac{m\Delta^{2}}{4\pi\hbar^{2}a},\end{equation}
 where the excitation energy $E_{\mathbf{k}}=[(\epsilon_{\mathbf{k}}-\mu)^{2}+\Delta^{2}]^{1/2}$
and the Fermi distribution function $f(x)=1/(1+e^{x/k_{B}T})$, we
obtain the full thermodynamic potential $\Omega=\Omega_{0}+\delta\Omega$.
All the observables are calculated straightforwardly following the
thermodynamic relations, once the chemical potential $\mu$ and the
order parameter $\Delta$ are determined. Number conservation in the
form $n=-\partial\Omega/\partial\mu$ is strictly satisfied, yielding
an exact identity for the pressure $P$ and energy density ${\cal E}$
of a unitarity gas: $P=2/3{\cal E}$~\cite{ho,thomas}. For simplicity,
in our calculations we determine the order parameter at the level
of mean field, using the gap equation $\partial\Omega_{0}/\partial\Delta=0$.
Part of our approach was also previously derived using a functional
integral method~\cite{engelbrecht}. In the case of the normal Fermi
liquid with vanishing order parameter, the usual NSR formalism is
recovered~\cite{nsr}.

This type of perturbation theory with \emph{bare} BCS Green functions
in the pair propagators constitutes the simplest description of the
BCS-BEC crossover, including the essential pair fluctuations. More
sophisticated approximations with dressed Green functions in the pair
propagators, \textit{i.e.}, the $GG_{0}$~\cite{chen} and the fully
self-consistent $GG$ schemes~\cite{sc94,lhpra}, have also been
proposed. In Ref.~\cite{hldpra}, we performed a comparative study
of these different approximation schemes for a unitarity gas in the
normal state. A related calculation in the superfluid phase has also
been carried out recently~\cite{sc06}. Compared with the latest
path-integral Monte Carlo simulations~\cite{bulgac,burovski}, our
analytic perturbation scheme seems to be the optimal choice for the
calculations of the type required for the entropy-energy relation.
However, there is a small region around the critical temperature where
none of the current calculations are reliable, and we see indications
of this in the data.

To include the effects of the trap, we employ the local density approximation
by assuming that the system can be treated as locally uniform, with
a position dependent local chemical potential $\mu\left(\mathbf{r}\right)=\mu-V(\mathbf{r})$,
where $V(\mathbf{r})$ is the trapping potential. The local entropy
and energy, calculated directly from the local thermodynamic potential
using thermodynamic relations, are then summed to give the total entropy
and energy. We note that in the presence of a harmonic trap, the exact
identity $P=2/3{\cal E}$ yields the virial theorem~\cite{thomas},
which states that the potential energy of the gas is a half of its
total energy.

\end{document}